%% file: wisec2020.tex
\documentclass[conference]{IEEEtran}

\IEEEoverridecommandlockouts
\usepackage{cite}
\usepackage{amsmath,amssymb,amsfonts}
\usepackage{algorithmic}
\usepackage{graphicx}
\usepackage{svg}
\usepackage{textcomp}
\usepackage{xcolor}
\usepackage{authblk}
\usepackage{amsmath}
\usepackage[export]{adjustbox}

\usepackage{paralist}

\usepackage{graphicx}

\usepackage{fancyhdr}
\begin{document}

\title{Investigating a Spectral Deception Loss Metric for Training Machine Learning-based Evasion Attacks}

\author{Matthew DelVecchio, Vanessa Arndorfer, William C. Headley}
\affil{Hume Center for National Security and Technology, Virginia Tech}
\affil{[matdd96, arndorvf, cheadley]@vt.edu}
\maketitle

\begin{abstract}

Adversarial evasion attacks have been very successful in causing poor performance in a wide variety of machine learning applications. One such application is radio frequency spectrum sensing. While evasion attacks have proven particularly successful in this area, they have done so at the detriment of the signal's intended purpose. More specifically, for real-world applications of interest, the resulting perturbed signal that is transmitted to evade an eavesdropper must not deviate far from the original signal, less the intended information is destroyed. Recent work by the authors and others has demonstrated an attack framework that allows for intelligent balancing between these conflicting goals of evasion and communication. However, while these methodologies consider creating adversarial signals that minimize communications degradation, they have been shown to do so at the expense of the spectral shape of the signal. This opens the adversarial signal up to defenses at the eavesdropper such as filtering, which could render the attack ineffective. To remedy this, this work introduces a new spectral deception loss metric that can be implemented during the training process to force the spectral shape to be more in-line with the original signal. As an initial proof of concept, a variety of methods are presented that provide a starting point for this proposed loss. Through performance analysis, it is shown that these techniques are effective in controlling the shape of the adversarial signal.

\end{abstract}



\section{Introduction} \label{sec:intro}
\input{sections/introduction.tex}

\begin{figure*}
\centering
    \includegraphics[width=0.85\linewidth]{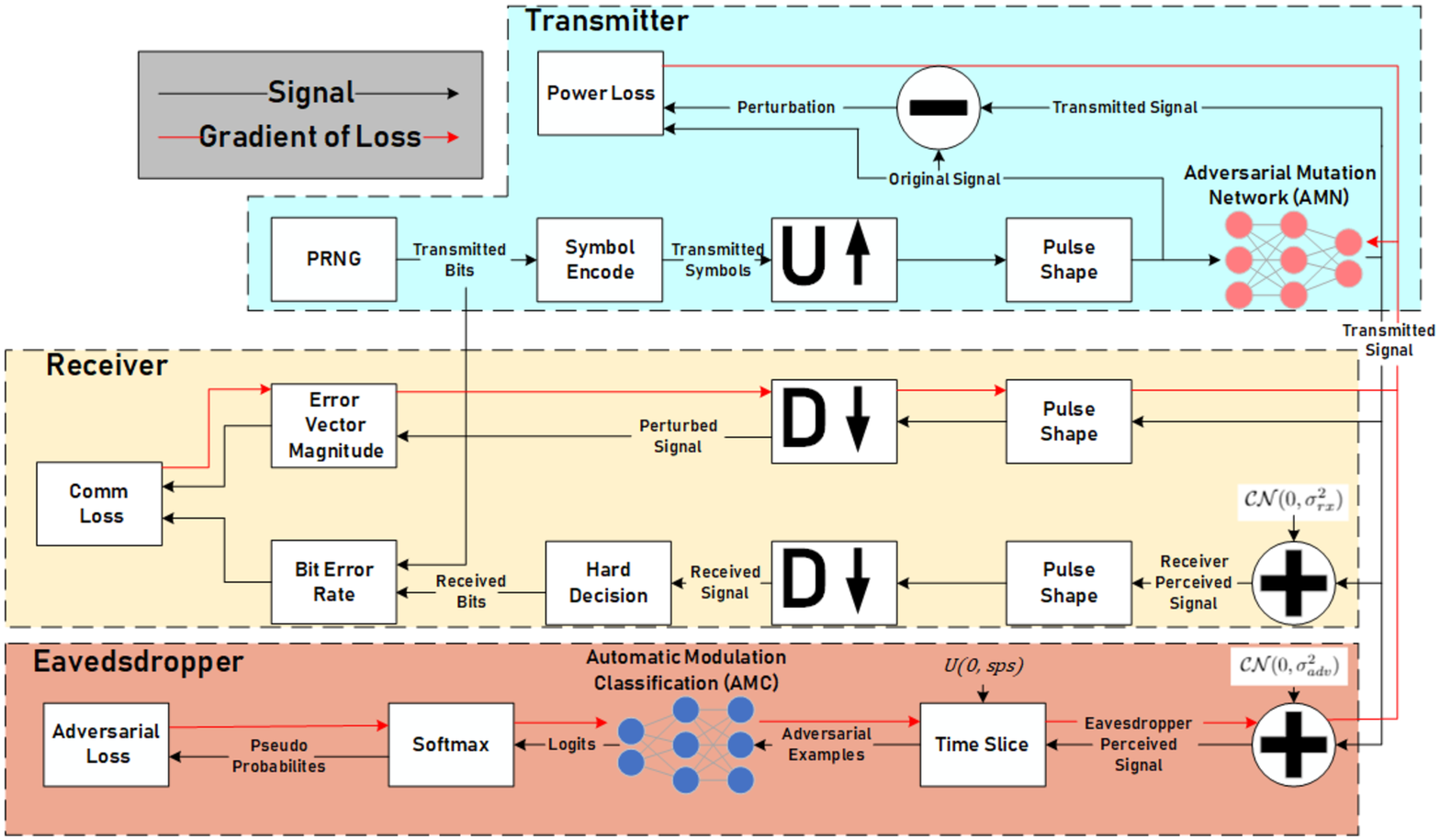}
    \caption{The training process that this work builds off of. It utilizes three losses, evasion loss, communication loss, and power loss each calculated at the eavesdropper, receiver, and transmitter respectively \cite{RN205}. This work will replace the power loss calculated in the transmitter, leaving the others untouched.}
    \vspace{-1em}
    \label{fig:background:training}
\end{figure*}

\section{Background} \label{sec:background}

\input{sections/background.tex}

\section{Spectral Deception Loss} \label{sec:spectrum}
\input{sections/spectrum.tex}

\section{Conclusion and Future Work} \label{sec:conclusion}

\input{sections/conclusion.tex}

\section{Acknowledgments}

This material is based upon work supported by the National Science Foundation under Grant Number 1303297. Any opinions, findings, and conclusions or recommendations expressed in this material are those of the author(s) and do not necessarily reflect the views of the National Science Foundation. This material was also aided as part of an undergraduate research program sponsored by the Naval Engineering Education Consortium and NSWC Crane. 


\bibliographystyle{IEEEtran}
\bibliography{man.bib}

\end{document}

%% file: sections/introduction.tex
Boosted by continued improvements in areas such as processing power, storage capacity, and architectural improvements, machine learning algorithms have seen increased usage in recent years and have shown great potential benefit in a wide variety of research fields. One such emerging research field is signal processing, where research has focused on utilizing recent advancements in machine learning to improve on traditional digital signal processing techniques through increased performance and/or a reduced need for \emph{a priori} knowledge. Example signal processing applications showing promise in their utilization of machine learning include spectrum signal detection, synthetic modulation schemes, direction of arrival calculation, jamming detection \cite{RN201, RN203}, and automatic modulation classifiers (AMC) \cite{RN9, RN10, RN12, RN202}, among many others. AMC research in particular has shown significant promise in utilizing machine learning to reduce requirements on pre-defined expert features by utilizing state-of-the-art convolutional neural networks for performing both feature extraction and classification tasks \cite{RN6, RN4}.

\begin{figure}
\centering
    \includegraphics[width=\linewidth]{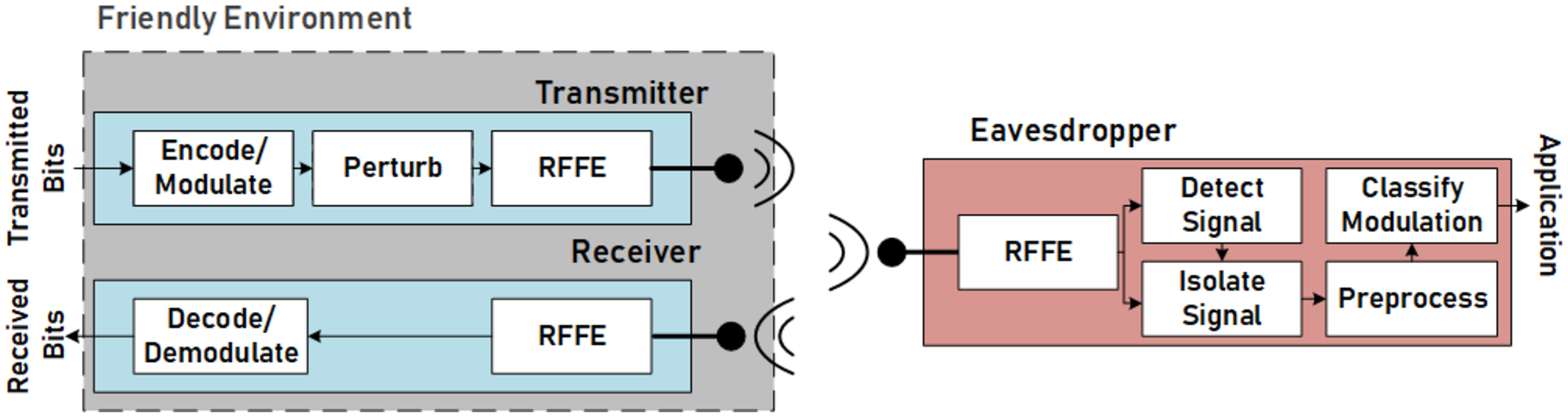}
    \caption{A wireless communications scenario in which an intended communications link is being eavesdropped by a machine learning based spectrum sensor. The transmitter utilizes an adversarial evasion attack to intelligently "perturb" its signal to evade the eavesdropper \cite{RN205}.}
    \vspace{-1em}
    \label{fig:background:ThreatModel}
\end{figure}

Given the improvements that machine learning can offer, and thus the adoption of such methods in real world applications, the security of these networks must be further considered. Recent research has shown that adversarial attacks can harm the performance of machine learning networks by forcing misclassifications or otherwise causing the network to operate in ways orthogonal to its intended use or desired application \cite{RN15, RN16, papernot}. Various adversarial techniques can be used to attack a machine learning network, such as poisoning \cite{RN22}, Trojan \cite{RN204}, and evasion attacks. In the context of attacking AMC networks, the focus of this work, evasion attacks have been used to make slight intelligent changes to signals so that a trained AMC machine learning algorithm misclassifies the signal \cite{RN1, RN26, RN30, RN207}. Therefore, these adversarial techniques can be used as a mitigation approach against eavesdroppers and other malicious actors.

When developing these attacks, there is a natural tradeoff that arises between security and the intended application. For example, while the goal of an evasion attack against an AMC machine learning algorithm is to cause a misclassification and/or reduce user confidence, it is important that the perturbed signal still accomplish its intended use of still being successfully received by its intended target. In the field of image recognition, this manifests in the idea that an image perturbed by an adversarial attack should still be easily discernible, and even viewed as untouched, by a human observing the image \cite{RN16}. 

The adversarial scenario considered in this work is illustrated by Figure 1. As previously mentioned, balancing the two conflicting goals of evasion of an eavesdropper and successful communication by an intended receiver is difficult and has been examined in previous work \cite{RN29, RN2, RN205, RN206}. In this work, the idea of successful perception of the signal at the receiver is driven using metrics such as bit error rate (BER) that indicate the success of the communication. In addition, this work presents a novel form of perception to be considered alongside BER, that of spectral integrity. The previous works in this area have shown that adversarial perturbations naturally tend to manifest out of the main lobe of the signal and lead to adversarial signals that do not hold well the same spectral shape as the original signal \cite{RN2}. This change in the spectral shape of the signal poses a problem to the success of the attack as the eavesdropper could leverage preprocessing stages to reduce the impact of the perturbation, such as with a filter, and potentially can lead to increased likelihood of detection that an attack is taking place since the spectral shape does not appear benign.

This work introduces a new loss metric for training machine learning based adversarial evasion attacks that helps maintain spectral integrity of the adversarially perturbed signal while still successfully achieving evasion and solid communication. Section II of this paper first provides a background on previous work done in this field and the particular evasion attack method used in this work. Section III introduces candidate spectral integrity loss metrics and provides relevant performance analysis. Finally, Section IV concludes this work and discusses future work based on these findings. 

%% file: sections/background.tex
Without proper care, evasion attacks used to fool an AMC machine learning algorithm generally have a drastic negative impact on the communication link between the transmitter and intended receiver. Recently, work by the authors and others have examined how these attacks can be improved in order to provide a better balance between these two conflicting goals. Hameed et. al. \cite{RN29} accomplished this by introducing a gradient descent training method to craft signal perturbations that utilize a combined target function that considers both evasion performance and BER. While BER is non-differentiable, and thus not suited to gradient based learning approaches, a gradient is estimated using simultaneous perturbation stochastic approximation (SPSA). This approaches offers improvement over previous methods where the perturbation was simply power limited in the hope that this would lead to decreased BER. Flowers et al. improved upon these prior works through the development of a so-called "communications-aware" attack \cite{RN2}. 

For the communications-aware attack, an Adversarial Residual Network (ARN) is leveraged in order to learn to make intelligent signal perturbations that balance the two opposing goals of evasion and communication. This approach utilizes three separate loss functions to accomplish this: adversarial loss, communication loss, and power loss. These three losses are each weighted and summed together to guide the ARNs learning process. The work of \cite{RN205} expanded on the communications aware architecture first introduced in \cite{RN2} in order to better utilize forward error correction (FEC), but it was found that the changes to the loss functions and transmitter architecture provided improvements beyond just utilizing FEC. The training framework presented in \cite{RN205}, and illustrated in Figure 2, serves as the foundation for the work presented in this paper.

As shown in Figure 2, the considered approach utilizes an Adversarial Mutation network (AMN) that is trained to create an intelligently perturbed signal given the original signal as input. This adversarial signal is what is transmitted to the intended receiver and intercepted by the eavesdropper. The AMN consists of a convolutional neural network (CNN). It is assumed here that the eavesdropper utilizes an AMC network with the architecture described in \cite{RN9} trained for BPSK, QPSK, 8PSK, 16QAM, and 64QAM. Each AMN is trained to create adversarial signals for just one modulation scheme at at a time. As previously mentioned, the AMN developed in \cite{RN205} utilizes three loss functions to train the AMN network, namely:

\begin{itemize}
    \item \emph{Adversarial Loss}: prioritizes the AMN's ability to successfully learn to avoid classification by the eavesdropper. It is calculated using the confidence of the eavesdropper in the true source modulation, $p_{s}$, determined using the output of the final softmax layer in the eavesdropper's AMC.
    \item \emph{Communication Loss}: prioritizes the AMN's ability to successfully learn to maintain the communication link between the transmitter and friendly receiver. It does this by using the calculated BER, $b_{r}$, as well as the error vector magnitude (EVM) between the clean symbols and the perturbed symbols, defined as $|S_{tx} - S_{tx+p}|$. BER is calculated using the original bits at the transmitter and the final bits decoded at the receiver after undergoing AWGN channel effects. The AWGN channel adds random noise between 0-20 dB. In this work it is assumed that the transmitter has access to the receiver in order to know the bits received. The BER is the true metric that the network wants to minimize, but is non-differentiable, so the EVM, which is differentiable, acts as a proxy for the BER and provides a gradient indicating the direction the weights should update. The BER then provides the magnitude of the update along this gradient. 
    \item \emph{Power Loss}: prioritizes the AMN's ability to learn to minimize the power of the perturbation so that the adversarial signal is close to the original signal. It does this by using the inverse of the signal-to-perturbation ratio (SPR).
\end{itemize}

During the AMN's training process, these three losses are each scaled and then summed together to create the total loss. These scaling factors allow for finer balancing between the communication and evasion goals. The scaling factors are $\alpha$ for adversarial loss, $\beta$ for communications loss, and $\gamma$ for power loss. More specifically, increasing a scaling factor relative to the others during training results in the corresponding loss being more highly prioritized. These loss constants are restricted such that they must sum to 1. Finally, the three loss functions are designed such that they all converge to zero. Therefore the network learns to minimize the loss values during training. Currently, these loss constants are estimated during training based on the rough needs of the system, such as whether evasion, communication, or power should be more important. A more exhaustive look into the best way to determine the values of the constants is left to future work. During the training process, the total loss is back-propagated through the CNN of the AMN to update the weights in order to create the most effective adversarial signal. The optimization technique used is Adam. To summarize, the loss functions are defined below:
\begin{equation}
    \mathcal{L}_{\text{total}} = \alpha \mathcal{L}_{\text{adv}} + \beta \mathcal{L}_{\text{comm}} + \gamma \mathcal{L}_{\text{pwr}}
    \label{eq:total_loss}
    \vspace{-0.75em}
\end{equation}

\begin{equation}
    \mathcal{L}_{\text{adv}} = -log(1-p_{s})
    \label{eq:l_adv}
    \vspace{-0.75em}
\end{equation}

\begin{equation}
    \mathcal{L}_{\text{comm}} = b_{r} \times EVM(S_{tx}, S_{tx+p})
    \label{eq:l_comm}
    \vspace{-0.75em}
\end{equation}

\begin{equation}
    \mathcal{L}_{\text{pwr}} = \frac{1}{E_{s}/E_{p}} = \frac{E_{p}}{E_{s}}
    \label{eq:l_pwr}
\end{equation}

 The architecture changes specified in \cite{RN205}, originally designed for use of forward error correction coding on the signal, allowed for improved spectral shape over the results seen in \cite{RN2}. This improvement was predominantly due to improvements in the power loss metric and the usage of AMN as opposed to an ARN.

In this work, the same framework described above is used. However, here the power loss is replaced with a novel loss metric termed spectral deception loss. The goal of this loss will be to more explicitly train the network to create adversarial signals that follow the same spectral shape as the original signal, while still balancing between the conflicting goals of evasion and intended communications. The rest of the architecture, including the adversarial and communication loss, remains unchanged. 

%% file: sections/spectrum.tex
In this section, a variety of candidate spectrum deception loss metrics are presented, and their different impacts on the adversarial signal's spectral content are analyzed, along with its performance on eavesdropper evasion and intended communication capabilities. As previously discussed, it is desirable for the adversarial signal to have a similar spectral shape as the original signal so that it avoids detection and defensive capabilities. In this work, we determine this similarity through the power spectral density (PSD) and associated phase plot of the original signal, perturbation, and combined adversarial signal. Due to space considerations, only the PSD and not the phase plots are shown as the PSD provides a much better indication of success. 

\subsection{Examining the Necessity of Deception Loss}
In the previous work, there was uncertainty over whether the power loss metric was sufficiently useful at providing the desired intent of maintaining the original shape of the signal.
This was due partially to the fact that the two main performance metrics, BER and evasion classification success, were driven directly by the communication and adversarial loss, respectively, and not by the power loss. Additionally, the power loss and communication loss were shown to push the network to converge in the exact same way for these metrics, as is shown in Figure 3, leading to unnecessary redundancy among these two losses.

\begin{figure}
\centering
    \includegraphics[width=0.92\linewidth]{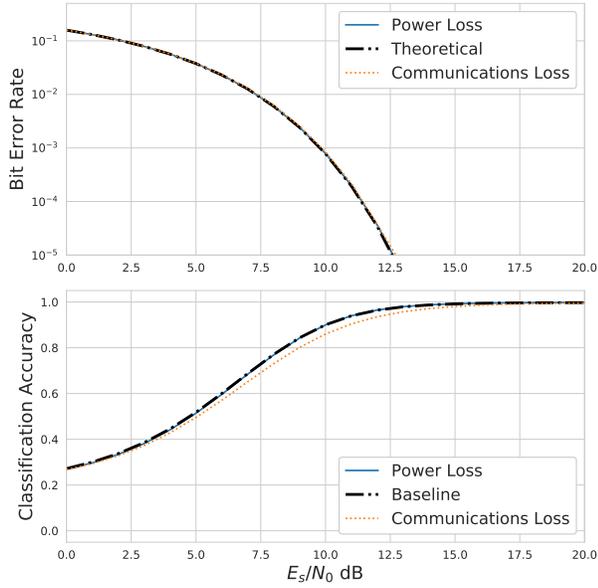}
    \caption{The BER and eavesdropper classification accuracy for QPSK adversarial signals when trained with either only communications loss or only power loss. The values are plotted over a range of 0-20 dB SNR. The theoretical values for the BER and classification accuracy of QPSK are shown.}
    \vspace{-1em}
    \label{fig:spectrum:comm_pow_ber_acc}
\end{figure}

It makes sense that these two losses would provide similar results for the chosen performance metrics. However, observation of the PSD of the resulting adversarial signal when prioritizing each loss highlights the true differences between them. An example of this difference is shown in Figure 4. Prioritizing the power loss results in a PSD shape for the perturbation that is similar to the original signal, only less powerful. On the other hand, prioritizing the communication loss results in a PSD that is more jagged in the center lobe and has significant side lobe content. From this result, it can be observed that the power loss metric steers the training of the AMN to keep the spectral shape of the original signal while the communication loss metric disregards the original signal shape as long as the intended receiver is minimally impacted.

While the power loss appears to provide the exact behavior desired to maintain spectral integrity, this is only true under an ideal scenario. In the power loss result shown in Figure 4, the power loss is the only loss prioritized. However, when being balanced with the communication and evasion losses, the shape, while still an improvement on previous work, no longer resembles a clean signal and has some side lobe content \cite{RN205}. Spectral deception loss is introduced as a solution to this problem so that the spectral integrity can be preserved even when successfully evading and communicating. The deception loss will operate in the frequency domain and thus allows for the AMN to better control the frequency content of the signal as opposed to the prior power loss metric that controls the time content of the signal. This should allow for better success in shaping the signal. As mentioned previously, this control over the spectral shape is desirable so that the attack can better avoid defenses such as filtering in the preprocessing stage of the eavesdropper. Previous work resulted in perturbations that had significant content in the side lobe. Such a perturbation could be weakened by a low pass filter that would remove this side lobe content and potentially render the attack ineffective. By forcing the perturbation to be more in lobe, the deception loss should help the attack remain robust to these forms of filtering and defense. 

\begin{figure}
\centering
    \includegraphics[width=0.92\linewidth]{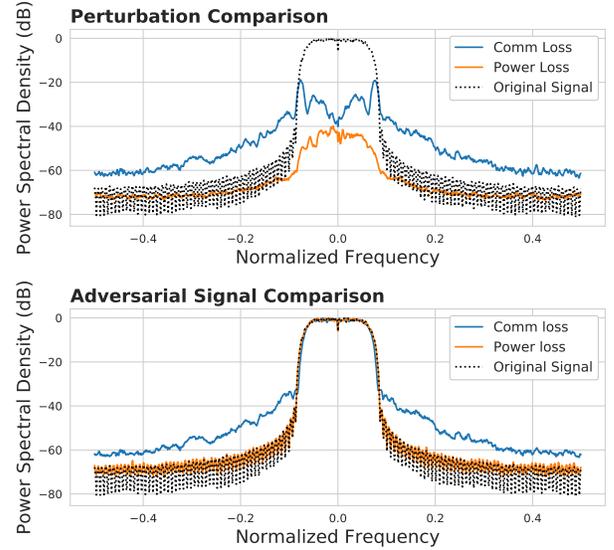}
    \caption{The PSD for both a perturbation created using only communication loss and one created using only power loss compared to the PSD of the original signal.}
    \vspace{-1em}
    \label{fig:spectrum:comm_pow_psd}
\end{figure}

\subsection{Deception Loss}

The deception loss method to be discussed within this work is based upon the frequency domain characteristics of the signal. More specifically, the proposed deception loss function operates on the Fast-Fourier Transform (FFT) of both the perturbation and original signal. This is done so that the perturbation lies more in-band and thus the adversarial signal will exhibit less side-lobe content and appear more benign. A function must be used in order to quantify the difference between the two FFTs. Two functions, Mean Squared Error (MSE) and Huber, are examined in this paper for their potential use in the deception loss.  

\subsubsection{MSE FFT Loss}

MSE is a regression loss function that determines the difference between expected and actual values. In this paper, MSE is used as the average squared difference between the FFTs of the original signal and the perturbation. MSE is defined as:
\begin{equation}
    \frac{1}{n}\sum_{i=1}^n (y_i - \hat{y}_i)^2
\end{equation}
where $y$ is the value of the original signal and $\hat{y}$ is the value of the perturbation. After calculation, the loss was normalized such that $0 \leq$ MSE $\leq 1$ to better align with the communication and adversarial loss values.

\subsubsection{Huber FFT Loss}

Although MSE is a good comparison metric for two functions, it is often heavily influenced by outliers. Huber loss mitigates the affect of outliers through an adjustable delta value, $\delta$. If the absolute difference between the expected and actual value is less than $\delta$, then Huber loss calculates their difference using an equation similar to MSE. Otherwise, the affect of the outlier is adjusted using the Mean Absolute Error (MAE) function. The Huber loss function is shown below.
\begin{equation}
    \begin{cases}
        \frac{1}{2} (y-\hat{y})^2 & |y-\hat{y}| \leq \delta,\\
        \delta|y-\hat{y}|-\frac{1}{2}\delta^2 & \text{otherwise}
    \end{cases}
\end{equation}
where $y$ is the value of the original signal and $\hat{y}$ is the value of the perturbed signal. Equation 6 specifies the function used to calculate the difference between two corresponding points in the FFTs of the original signal and perturbation. These differences are then summed and divided by the total number of points to obtain an average, like what is done with MSE. The value of $\delta$ used in this work is 1. Similar to MSE, Huber loss is normalized such that the loss value is contained between 0 and 1.

\subsection{Results}
The primary qualitative metric used when examining the success of the various spectral deception loss methods at maintaining the spectral shape was visual inspection of the PSD. Quantitative metrics used to validate the success of the considered metrics include the BER of the intended communications link and the achieved reduction in classification accuracy of the eavesdropper. The results presented in this section are predominantly examined with AMNs trained for QPSK modulated signals. However, other modulation schemes were also tested and exhibited the same characteristics. The eavesdropper's AMC used in this work was trained on BPSK, QPSK, 8PSK, 16QAM, and 64QAM.

\begin{figure}
\centering
    \includegraphics[width=0.94\linewidth]{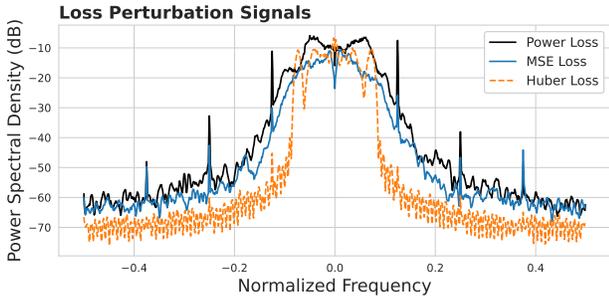}
    \caption{The PSD for the perturbations created by the original power loss and both the MSE and Huber loss methods for the FFT-based approach on BPSK signals.}
    \vspace{-1em}
    \label{fig:spectrum:fft_pert_psd}
\end{figure}

\begin{figure}
\centering
    \includegraphics[width=0.94\linewidth]{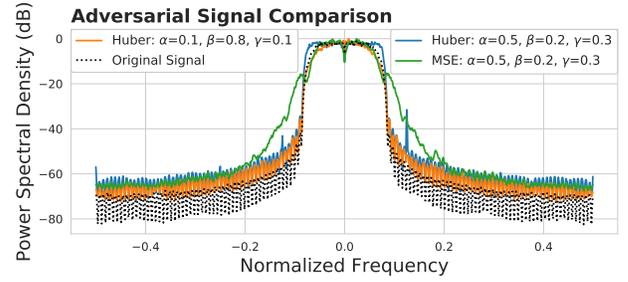}
    \caption{The PSD for the adversarial signals created by the MSE and Huber loss methods for the FFT-based approach on QPSK signals.}
    \vspace{-1em}
    \label{fig:spectrum:fft_total_psd}
\end{figure}

As mentioned previously, this FFT-based approach was tested using both the MSE loss function and the Huber loss function. Figure 5 shows the resulting PSDs of just the perturbation for the MSE loss, Huber loss, and the original power loss from the prior work. This figure illustrates that there is slight improvement with the MSE method over the power loss metric, but very minimal. However, the Huber loss method exhibits much better behavior over the power loss metric given that the shape of the perturbation is much more in-band to the original signal. This difference is due to the fact that the Huber loss is able to better handle situations of extreme error, which can occur during the training process especially at the start of training. Figure 6 shows the PSDs of the resulting adversarial signals. As can be seen from this figure, there is a trade off between the MSE method and the Huber method with respect to side lobe growth vs. main lobe corruption.

\begin{figure}
\centering
    \includegraphics[width=0.94\linewidth]{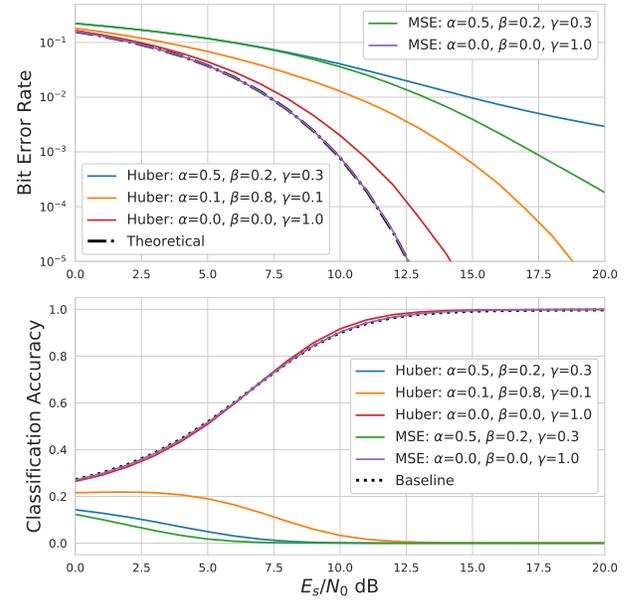}
    \caption{The BER and eavesdropper classification accuracy for QPSK adversarial signals when with the deception loss that is done on the FFT using both Huber and MSE loss. The signals correspond to those shown in Figure 6.}
    \vspace{-1em}
    \label{fig:spectrum:fft_ber_acc}
\end{figure}

As expected, this trade-off in spectral shape performance comes at the detriment of intended communication performance. Figure 7 shows the BER and eavesdropper classification accuracy over the SNR range of 0-20 dB of the two methods, along with the theoretical QPSK bit error rate with no perturbation added. When using loss constants of $\alpha=0.5$, $\beta=0.2$, and $\gamma=0.3$, the BER rate for the Huber method is much worse than that of the MSE method. Additionally, when the MSE deception loss is the only loss considered during training (i.e. $\alpha$ and $\beta$ are set to 0), the BER converges to the theoretical curve, which does not occur for the Huber loss. However, by adjusting the loss constants, the communication performance can be made better as is shown by the Huber result with $\alpha=0.1$, $\beta=0.8$, and $\gamma=0.1$. Naturally, this does lead to worse evasion performance. Interestingly, in Figure 6 it can be observed that the resulting spectral shape of the adversarial signal does not seem to drastically change for this second Huber trial even though the deception loss is less prioritized. This shows that the constants can be adjusted to meet the needs of the attack and that when using Huber loss in the deception loss, the spectral shape can be maintained even when less prioritized (so more priority can be spent on evasion or communication improvement).

%% file: sections/conclusion.tex
The results of this work show the benefit of utilizing a spectral deception loss metric within the considered machine learning based adversarial evasion attack. The considered FFT-based methods of developing this metric provided solid improvements over the prior work and can be used as the foundation for future work. Utilizing the FFT, two loss metrics were considered, namely MSE and Huber losses. Performance analysis demonstrated that the Huber loss was more successful at maintaining the spectral shape of the original signal over the MSE loss, at the cost of decreased intended communication performance.

While these results show promise, there is still much future work to develop the concept further. The various deception loss methods presented in this work are intended as starting points and improving upon these may offer greater success. For example, one simple adaptation could come in the form of completing a more exhaustive parameter search over the configurations for the deception loss. Additionally, other functions than the FFT investigated here could be used to determine and quantify the difference between the original signal and the adversarial signal. For example, minimizing the difference between the resulting PSDs and associated phases could be examined. Finally, while mean squared error (MSE) and Huber are good for determining the difference between corresponding elements in an array of data, such as with time domain samples, they may not be the most appropriate for the frequency domain. Other functions, such as Fr\'echet distance, may provide better comparisons of similarity and should be further studied. 

The predominant method used in this work to determine success of the loss was to qualitatively observe if the perturbation was concentrated in the main lobe of the signal. While this may be sufficient in determining whether a human operator can detect the adversarial signal, future work should examine whether this adapted attack framework would be effective in evading detection by a machine learning algorithm aimed at detecting these attacks. Additionally, previous work has assumed oversampling of the signal by the eavesdropper which provides a larger attack vector for the evasion attack in terms of available bandwidth outside of the signal's main lobe. Future work should loosen this assumption in order to better test the success of the deception loss. Recent work has focused on strategies that make the classifier networks more robust against attacks such as utilizing curriculum training \cite{RN208}. Future work should examine the success of evasion attacks against such defensive techniques when employing the deception loss.

While the concept of a spectral deception loss is an extremely new area of focus, this work has shown that it is one that offers great potential in the effort to mask the limitations and distinguishing characteristics of existing evasion attacks. 